\documentclass[aps,prb,preprint,superscriptaddress,showpacs,showkeywords]{revtex4}
\usepackage{amssymb}
\usepackage{graphicx}% Include figure files
\usepackage{dcolumn}% Align table columns on decimal point
\usepackage{bm}% bold math
\usepackage{amsmath}
\usepackage{latexsym}
\usepackage{epsfig}
\usepackage{amsbsy}
\usepackage{array}
\usepackage{setspace}
\usepackage{multirow}
\usepackage{float}
\usepackage{threeparttable}
\begin{document}

\title{Compression and phase diagram of lithium hydrides at elevated
pressures and temperatures by first-principles calculations}

\author{Yang M. Chen}
\affiliation{The Institute of Atomic and Molecular Physics, College of Physical Science and Technology, Sichuan University, Chengdu, Sichuan, People{\textquoteright}s Republic of China,610065}
\affiliation{National Key Laboratory of Shock Wave and Detonation Physics,
Institute of Fluid Physics, P.O. Box 919-102, Mianyang, Sichuan 621900, China}
\author{Xiang R. Chen}
\email{xrchen@scu.edu.cn}
\affiliation{The Institute of Atomic and Molecular Physics, College of Physical Science and Technology, Sichuan University, Chengdu, Sichuan, People{\textquoteright}s Republic of China,610065}
\author{Qiang Wu}
\email{wuqianglsd@163.com}
\affiliation{National Key Laboratory of Shock Wave and Detonation Physics,
Institute of Fluid Physics, P.O. Box 919-102, Mianyang, Sichuan 621900, China}
\author{Hua Y. Geng}
\email{s102genghy@caep.cn}
\affiliation{National Key Laboratory of Shock Wave and Detonation Physics,
Institute of Fluid Physics, P.O. Box 919-102, Mianyang, Sichuan 621900, China}
\author{Xiao Z. Yan}
\affiliation{The Institute of Atomic and Molecular Physics, College of Physical Science and Technology, Sichuan University, Chengdu, Sichuan, People{\textquoteright}s Republic of China,610065}
\affiliation{National Key Laboratory of Shock Wave and Detonation Physics,
Institute of Fluid Physics, P.O. Box 919-102, Mianyang, Sichuan 621900, China}
\author{Yi X. Wang}
\affiliation{The Institute of Atomic and Molecular Physics, College of Physical Science and Technology, Sichuan University, Chengdu, Sichuan, People{\textquoteright}s Republic of China,610065}
\affiliation{National Key Laboratory of Shock Wave and Detonation Physics,
Institute of Fluid Physics, P.O. Box 919-102, Mianyang, Sichuan 621900, China}
\author{Zi W. Wang}
\affiliation{National Key Laboratory of Shock Wave and Detonation Physics,
Institute of Fluid Physics, P.O. Box 919-102, Mianyang, Sichuan 621900, China}

\date{\today}

\begin{abstract}
High pressure and high temperature properties of AB (A = $^6$Li, $^7$Li; B = H, D, T) are investigated with first-principles method comprehensively. It is found that the H$^{-}$ sublattice features in the low-pressure electronic structure near the Fermi level of LiH are shifted to that dominated by the Li$^{+}$ sublattice in compression. The lattice dynamics is studied in quasi-harmonic approximation, from which the phonon contribution to the free energy and the isotopic effects are accurately modelled with the aid of a parameterized double-Debye model. The obtained equation of state (EOS) matches perfectly with available static experimental data. The calculated principal Hugoniot is also in accordance with that derived from shock wave experiments. Using the calculated principal Hugoniot and the previous theoretical melting curve, we predict a shock melting point at 56 GPa and 1923 K. In order to establish the phase diagram for LiH, the phase boundaries between the B1 and B2 solid phases are explored. The B1-B2-liquid triple point is determined at about 241 GPa and 2413 K. The remarkable shift in the phase boundaries by isotopic effect and temperature reveal the significant role played by lattice vibrations. Furthermore, the Hugoniot of the static-dynamic coupling compression is assessed. Our EOS suggests that a precompression of the sample to 50 GPa will allow the shock Hugoniot passing through the triple point and entering the B2 solid phase. This transition leads to a discontinuity with 4.6\% volume collapse, about four times greater than the same B1-B2 transition at zero temperature.
\end{abstract}

\keywords{lithium hydride, equation of state, high pressure, first-principles, phase diagram}
\pacs{63.20.dk, 64.60.A-, 64.70.D-}

\maketitle

\section{INTRODUCTION}
As the lightest ionic compound, as well as the highest mass content of hydrogen and the highest melting
point of 965 K at ambient pressure \cite{Messer543} in alkali metal hydrides, LiH has been widely studied and applied
in the fields of hydrogen storage \cite{George5454}, thermonuclear fusion, and aviation and space industries \cite{Van434, Tyutyunnik539, Bradtke20, Mueller633}. Early static compression experiment using diamond anvil cell (DAC) showed that LiH occupies an FCC lattice and orders in NaCl (B1) structure at ambient condition, and this structure is maintained up to at least 36 GPa (96 GPa for LiD) \cite{Loubeyre10403}. Under this pressure, all other alkali hydrides were observed to transform into CsCl (B2) phase (NaH at 29.3 GPa, KH at 4.0 GPa, RbH at 2.2 GPa, CsH at 0.83 GPa) \cite{Duclos7664, Hochheimer139, Ghandehari2264}. However, the same
structural transition in LiH has yet to be observed, which stimulates broad and continuous high pressure
experimental and theoretical researches. Recently, by analyzing the x-ray diffraction (XRD) data obtained in
DAC experiment \cite{Lazicki054103}, it was shown that at room temperature LiH remains in the B1 structure
under pressures up to 252 GPa, the highest pressure having been studied experimentally so far. In particular,
the diffraction and Raman data indicated that the B1-B2 phase transition, as well as the accompanied
metallization, may not be far beyond 252 GPa \cite{Lazicki054103}.

With theoretical methods, the pressure-induced B1-B2 structural transition and the insulator-metal transition in LiH at low temperatures \cite{Hammerberg617, Martins7883, Zhang104115, Mukherjee103515, Wang470, Lebegue562, Zurek17640, Yu086209, Hama348} have been extensively investigated. The mechanism of the B1-B2 structural transition \cite{Zhang104115, xie2008origin, Mukherjee103515} is often interpreted using phonon softening and elastic instability. On the other hand, the insulator-metal transition was shown to occur prior to the B1-B2 transition by both the local density approximation (LDA) and semi-local generalized gradient approximation (GGA). This might be due to that LDA and GGA usually tend to underestimate the energy gap. By using all-electron GW approximation,
L\`{e}begue $et$ $al.$ \cite{Wang470} argued that the structural transition and metallization in LiH
should occur simultaneously at a pressure of 329 GPa. This transition pressure is close to the 313 GPa and
327 GPa calculated by Wang $et$ $al.$ \cite{Lebegue562} and Mukherjee $et$ $al,$ \cite{Mukherjee103515} respectively.
They employed the GGA and full-potential linearized augmented plane wave (LAPW) method as implemented in WIEN2K package.
Zurek $et$ $al.$ \cite{Zurek17640} also reported a transition pressure of 360 GPa calculated by VASP with the projector augmented
wave (PAW) pseudopotential and the Perdew-Burke-Enzerhof (PBE) approximation of the density-functional theory (DFT).

It should be noted that all of the above-mentioned calculations did not take the zero-point energy (ZPE) into account.
If including the ZPE of harmonic phonons at the level of GGA and density-functional perturbation theory (DFPT), the B1-B2 phase transition pressure was predicted to be 308 GPa \cite{Yu086209}. Using the plane-wave pseudopotential approach within the framework of DFT and DFPT, Zhang $et$ $al.$ \cite{Zhang976} discussed the electronic, lattice dynamic, and thermodynamic properties of AB (A = $^6$Li, $^7$Li; B = H, D, T) at ambient conditions. They confirmed that the lightest isotope $^6$LiH has the largest zero-point motion in a harmonic approximation. This implies that the isotopic effect may play an important role in the B1-B2 phase transition of lithium hydrides. The isotopic shift in the equation of state (EOS) between $^7$LiH and $^7$LiD at pressures up to 45 GPa was measured using DAC experiment \cite{Loubeyre10403}. The DFPT calculation \cite{Roma203} (only up to 10 GPa) showed that the isotopic shift is mainly due to the difference in ZPE. At pressures up to 20 GPa, Dammak $et$ $al.$ illustrated the anharmonic contribution to the lattice vibrations and to the isotopic pressure shift between $^7$LiH and $^7$LiD by using the quantum thermal bath molecular dynamics (QTB-MD) within DFT-GGA calculations \cite{Dammak435402}. Even with these extensive studies, the agreement with experimental data did not improve systematically, and the contribution of the lattice vibrations and the isotopic shift in lithium hydrides is still controversial. Furthermore, there are very few researches dedicated to the finite temperature behavior of lithium hydrides, and the finite temperature phase diagram is almost uncharted.

In this paper, the vibrational spectrum of lithium hydrides at elevated pressures are accurately determined by $ab$ $initio$ quasiharmonic calculations, from which the equation of state (EOS) and the phase diagram are derived with the aid of a parameterized double-Debye model. The shock compression behavior and the change in the shock path by precompression are also assessed. The theoretical and computational details are described in the next section. In section III, the results and discussions are presented.
The paper is summarized with conclusions given in section IV.

\section{COMPUTATIONAL METHODOLOGY}

Generally, the thermodynamics and the finite pressure-temperature phase diagram of a substance is
determined by the Gibbs free energy, which consists of three parts in a solid:
(i) the cold energy at zero temperature with nuclei at their equilibrium positions,
(ii) the vibrational free energy contributed from lattice dynamics, and (iii) the free energy of thermal electrons \cite{Chisolm104103}. In computer simulations, especially with first-principles total energy calculations, one usually obtains a set of discrete data of energy versus atomic volume, rather than a continuous curve of energy as an analytic function of density. To facilitate the practical application or post processing of the data, one would prefer to fit the discrete data to an analytical function or an equation of state, and then to derive a continuous and smooth curve. This not only endows the numerical data with physical implications, but also extends their application range greatly, if an adequate EOS model has been used. In this work, we fit the $ab$ $initio$  cold energies of the candidate solid phases to the Vinet EOS \cite{vinet1989universal}
\begin{equation}\label{EcV}
{{E}_{c}}(V)={{E}_{0}}+\frac{4{{V}_{0}}{{K}_{0}}}{{{K}_{m}}^{2}}[1-(1-\frac{3}{2}\eta
K_{m}^{{}})\exp (\frac{3}{2}\eta K_{m}^{{}})],
\end{equation}
in which
\begin{equation}\label{eta}
\eta =[1-{{(\frac{V}{{{V}_{0}}})}^{1/3}}]\text{ , }{{K}_{m}}={{{K}'}_{0}}-1\text{ , }
{{K}_{0}}=-V{{(\frac{\partial P}{\partial V})}_{0}}\text{ },\text{ }{{{K}'}_{0}}=\frac{\partial
{{K}_{0}}}{\partial P}.
\end{equation}
Here $V_0$, $K_0$, and $K^{'}{_0}$ are the specific volume, the bulk modulus, and
the derivative of bulk modulus with respect to pressure at the given reference state, respectively.
The cold pressure is then evaluated by ${{P}_{c}}=-\frac{\partial {{E}_{c}}}{\partial V}$. Please note that the Vinet EOS is
appropriate for LiH by comparison with other EOS models such as Murnaghan \cite{murnaghan1937finite}, Birch-Murnaghan \cite{birch1947finite} and Natural strain \cite{poirier1998logarithmic}. (Table SI in the supplemental material compares these EOS models with the experimental data)

The vibrational free energy can be modelled by semiempirical models such as Einstein or Debye model. Since the
parameters in these models usually are determined according to experiments performed at ambient conditions, their applicability
to high pressures is restricted. Alternatively, the phonon spectra can be calculated directly using first-principles quasi-harmonic approximation (QHA). This approach does not rely on any empirical input, and has high accuracy and unlimited application range (in principle it can be applied as long as the solid phase is dynamically stable).

In the quasi-harmonic approximation, the vibrations are treated as a gas of 3$N$ non-interacting phonons with frequencies ${\omega _{i{\rm{ }}}}$ depending on the atomic volume, where $N$ is the number of atoms per primitive cell. The vibrational free energy $F_{FP}$ in QHA is expressed as
\begin{equation}\label{Fvib}
F_{FP}^{} = \sum\limits_{j = 1}^{3N} {\left[ {\frac{{\hbar {\omega _j}}}{2} + {k_B}T\ln (1 - e^{-\hbar {\omega _j} /{k_B}T})} \right]},
\end{equation}
where ${k _B}$ is the Boltzmann's constant, and $T$ is the temperature. The thermal pressure is given by
\begin{equation}\label{pth}
P_{th}^{} =  - {\left( {\frac{{\partial {F_{FP}}}}{{\partial V}}} \right)_T} = \sum\limits_{j = 1}^{3N} {\left[ {\frac{{\hbar {\omega _j}{\gamma _j}}}{{2V}} + \frac{{{{\hbar {\omega _j}{\gamma _j}} \mathord{\left/
 {\vphantom {{\hbar {\omega _j}{\gamma _j}} V}} \right.
 \kern-\nulldelimiterspace} V}}}{{{e^{-\hbar {\omega _j} /{k_B}T}} - 1}}} \right]},
\end{equation}
where the mode Gr\"{u}neisen ratio
${\gamma _j} =  - {{\partial \ln {\omega _j}} \mathord{\left/
 {\vphantom {{\partial \ln {\omega _j}} {\partial \ln V}}} \right.
 \kern-\nulldelimiterspace} {\partial \ln V}}$ has been introduced. In practice, it is difficult to compute ${\gamma _j}$.
Alternatively, the vibrational free energy can be formulated as

\begin{equation}\label{FvibQHA}
F_{FP}^{} = \int\limits_0^\infty  {\left[ {\frac{{\hbar \omega }}{2} + {k_B}T\ln (1 - {e^{-\hbar \omega /{k_B}T}})} \right]} {g_{FP}}(\omega )d\omega
\end{equation}
by using the phonon density of states (phDOS) $g_{FP}(\omega)$. It is evident that $F_{FP}^{}$ is completely determined by $g_{FP}(\omega)$.

The phDOS usually can be evaluated only on a discrete grid of volume. Therefore, direct application of QHA is limited. Especially, a very fine grid is required if one wishes to obtain an accurate thermal pressure from Eq. (4). Analogous to the first-principles cold energy $E_c$, it is desirable to represent the QHA results by an analytic model. A good model has the capability to both interpolate and extrapolate the discrete QHA data, thus only a few QHA caculations are required to derive the accurate and wide range thermodynamics. In addition to this benefit in computational efficiency, an accurate EOS model with fewer parameters is good to integrate into hydrodynamics code for macroscopic simulations. Furthermore, numeric values of QHA free energy inevitably contain artificial noise arisen from computation precision. This noisy fluctuation is vital when calculating the phase boundaries from the intersection of free energies. Fitting the QHA data to a model can remove these fluctuations effectively. Different from cold energy, there are very few thermal EOS available for lattice vibrations. For an ionic compound such as LiH, the simple Debye model incorrectly treats the optical branches as acoustic modes. In this work, we will employ an improved variant of Debye model, i.e., the double-Debye model, to tackle this problem. Parameters of this model are determined by fitting to first-principles QHA phonon spectra. As will be shown below, this double-Debye model accurately reproduces the free energy of QHA, and is a faithful representation of the latter.

In the double-Debye model, the total phDOS $g(\omega)$ is given by a linear combination of the density of states of two standard single-Debye model (1DM) (here we use ``single'' to emphasize that it has just one
Debye temperature), which is
\begin{equation}\label{gdomega}
{{g}_{D}}(\omega )={{\xi }^{A}}g_{D}^{A}(\omega )+{{\xi }^{B}}g_{D}^{B}(\omega ).
\end{equation}
Here $g_{D}^{A(B)}(\omega )$ is the standard DOS of a Debye model, and has nonzero value of
$\frac{3{{\omega }^{2}}}{\left(\omega _{D}^{A(B)}\right)^3}$ only when $\omega \text{ }\le
\text{ }\omega _{D}^{A(B)}$, where ${{\omega }_{D}}$ is the corresponding Debye frequency
that relates to the Debye temperature by ${{k}_{B}}{{\theta }_{\text{D}}}\text{ }=\text{ }\hbar \omega _{D}^{{}}$.
The double-Debye is devised to reproduce the ZPE of QHA as $T$ $\rightarrow$ 0 K exactly and the
high-temperature expansion of the harmonic free energy up to the 2nd order, all of them are dictated
by the first-principles QHA phDOS ${{g}_{FP}}(\omega)$. These lead to three constraints on
the phonon characteristic temperatures $\theta_0$, $\theta_1$, and  $\theta_2$ as \cite{Chisolm104103}:
\begin{equation}\label{kB0}
{{k}_{B}}{{\theta }_{0}}=\hbar {{e}^{(1/3)}}\exp \left( \int{\ln (\omega ){{g}_{FP}}(\omega )d\omega }) \right),
\end{equation}
\begin{equation}\label{kB1}
{{k}_{B}}{{\theta }_{1}}=\frac{4}{3}\int{\hbar \omega {{g}_{FP}}(\omega )d\omega },
\end{equation}
\begin{equation}\label{kB2}
{{k}_{B}}{{\theta }_{2}}={{\left( \frac{5}{3}\int{{{(\hbar \omega )}^{2}}{{g}_{FP}}(\omega )d\omega } \right)}^{1/2}}.
\end{equation}
By using $\theta_0$, $\theta_1$, and $\theta_2$, the Debye temperatures $\theta_A$  and
$\theta_B$ ($\theta_A$ $\leq$ $\theta_B$) (which give rise to the respective density of state $g_{D}^{A}(\omega)$
and $g_{D}^{B}(\omega)$) must satisfy a set of nonlinear equations:
\begin{equation}\label{1}
1={{\xi }^{A}}+{{\xi }^{B}},
\end{equation}
\begin{equation}\label{lntheta}
\text{ln}\left( {{\theta }_{0}} \right)={{\xi }^{A}}\text{ln}\left( {{\theta }_{A}}
\right)+{{\xi }^{B}}\text{ln}\left( {{\theta }_{B}} \right),
\end{equation}
\begin{equation}\label{theta}
{{\theta }_{\text{1}}}={{\xi }^{A}}{{\theta }_{A}}+{{\xi }^{B}}{{\theta }_{B}},
\end{equation}
\begin{equation}\label{theta2}
{{\theta }_{2}}_{{}}^{2}={{\xi }^{A}}\theta _{A}^{\;2}+{{\xi }^{B}}\theta _{B}^{\;2}\text{ }.
\end{equation}

Solving these equations gives the solution for ${{\xi}^{A}}$, ${{\xi}^{B}}$, $\theta_A$, and $\theta_B$,
which then determine the double-Debye model by Eq. (\ref{gdomega}). It is worth noting that all of these
parameters are a function of the specific volume. The obtained phonon DOS $g_{D}^{{}}(\omega )$,
though has features only qualitatively similar to the original ${{g}_{FP}}(\omega )$, can reproduce the
vibrational free energy very accurately. In order to account for the variation of the phonon DOS with
respect to compression, the Gr\"{u}neisen parameters ${{\gamma }_{\{0,A,B\}}}$ are introduced and defined as
\begin{equation}\label{lntheta0AB}
-\frac{d\text{ }\ln{{\theta }_{\{\text{}0,A,B\}}}}{d\text{ }\ln V}\equiv\text{ }
{{\gamma }_{\{0,A,B\}}}={{\alpha }_{\{0,A,B\}}}\text{ }+\text{ }{{\beta }_{\{0,A,B\}}}V.
\end{equation}
The solution of Eq. (14) is
\begin{equation}\label{theta0AB}
{\theta _{\{ 0,A,B\} }}(V) = {\rm{ }}\theta _{\left\{ {0,A,B} \right\}}^0{(\frac{V}{{{V_{ref}}}})^{ - {\alpha _{\{ 0,A,B\} }}}}{\rm{exp }}\left[ {{\beta _{\{ 0,A,B\} }}({V_{ref}} - V)} \right],
\end{equation}
where $\theta _{\left\{ {0,A,B} \right\}}^0$ is the value of ${\theta _{\{ 0,A,B\} }}$ at the reference state with a volume of ${V_{ref}}$. In this way, the whole QHA free energy over a wide pressure and temperature range can be represented by a simple model with only nine parameters: $\theta _{\left\{ {0,A,B} \right\}}^0$, ${\alpha _{\{ 0,A,B\} }}$, and ${\beta _{\{ 0,A,B\} }}$.

Finally, the phonon contribution to the total free energy is expressed as:
\begin{equation}\label{FFP}
{{F}_{FP}}(V,T)\approx {{F}_{D}}(V,T)={{\xi }^{A}}{{F}_{A}}(V,T)+{{\xi }^{B}}{{F}_{B}}(V,T),
\end{equation}
with
\begin{equation}\label{FAB}
{{F}_{A(B)}}(V,T)={{k}_{B}}T\left\{ \frac{9{{\theta }_{A(B)}}}{8T}+3\ln \left[
1-{{e}^{-\frac{{{\theta }_{A(B)}}}{T}}} \right]-D(\frac{{{\theta }_{A(B)}}}{T}) \right\},
\end{equation}
in which the Debye function is given by
\begin{equation}\label{Dy}
D(y)=\frac{3}{{{y}^{3}}}\int_{0}^{y}{\frac{{{x}^{3}}}{\exp (x)-1}}dx.
\end{equation}

In this work, the cold energy is calculated with DFT \cite{HohenbergB864, KohnA1133} and plane-wave pseudopotential method,
as implemented in the Vienna Ab-initio Simulation Package (VASP) \cite{Kresse11169, Kresse15}. The Perdew-Burke-Ernzerhof (PBE)
parameterization of the electronic exchange-correlation energy functional \cite{Perdew3865}
is used. The interaction between ions and valence electrons is described by the projector
augmented-wave (PAW) pseudopotentials\cite{Blochl17953, Kresse1758}.
The kinetic energy cutoff for the plane-wave basis set is taken as 900 eV, a 25 $\times$ 25 $\times$ 25
Monkhorst-Pack grid for the k-points sampling is used for both B1 and B2 structures.
The convergence of these parameters is well checked, with the uncertainty in the total energy
less than 1 meV per atom.

If ignore the effects of electron-phonon (E-P) coupling on electronic structure, the electronic structures of all isotopes are the same in $ab$ $initio$ calculations. Since the major contribution to isotopic effects comes from lattice dynamics, we imposed this approximation in our work. In this case we alter the atomic mass in the standard pseudopotential to obtain the isotopic dynamics, which is a common practice when studying isotopic effects. Lattice dynamics and phonon density of states are calculated by using PHONOPY package\cite{Togo134106}, in which the force constants are approximated with the small displacement
method. The required forces are evaluated using VASP, with a supercell containing 128 atoms
in B1 phase and 250 atoms in B2 phase, respectively. A 4 $\times$ 4 $\times$ 4 k-point mesh is
used to sample the first Brillouin zone. The plane wave basis set cutoff is increased to 1000 eV.
The convergence of the obtained forces is carefully checked to ensure that the uncertainty in the ZPE is
less than 1 meV per atom. The phDOS are evaluated on a discrete volume grid. They are then fitted to the
double-Debye model (2DM) as briefed above, which has been successfully applied to calculate the phonon
free energy of dense hydrogen \cite{Caillabet094101} and carbon \cite{Correa024101}. For the purpose of comparison, the single-Debye model (1DM) is also evaluated in this work.

It should be noted that the contribution of thermoelectrons to the free energy in lithium
hydrides is very small within our considered pressure and temperature range, and thus is neglected.

\section{RESULTS AND DISCUSSION}
\subsection{Electronic structures}
LiH is a large gap insulator at ambient condition. The direct energy gap is about 4.94 eV by
the reflectance measurement \cite{Kondo367}. Our calculated band gap at zero pressure with LDA
and GGA is 2.65 and 2.95 eV, respectively. As other calculations reported in the literature\cite{Lebegue562, Zurek17640},
this underestimates the band gap. Using GW approximation \cite{Shishkin035101, Shishkin235102, Fuchs115109, Shishkin246403},
we obtained a band gap of 4.80 eV, slightly smaller than the experimental value. It was reported
that a simple self-interaction correction \cite{Baroni4077} is able to generate a band gap of 4.93 eV,
in perfect agreement with the experiment result. Because of the good performance of GW method for the band gap,
all electronic structure calculations described below were carried out using this method.

In order to understand the electronic structure, the wave function is usually decomposed
by projection onto atom-centered spherical orbitals with different angular momentum,
and then to construct the differential charge density with respect to the atomic superposition to analyse the chemical
bonding and charge transfer between atoms. It is a powerful tool to understand how quantum nature of electrons dictates material properties. Alternatively, for ionic compound such as LiH, since one expects a complete charge transfer
from Li-$2s$ to H-$1s$, it could be regarded as a pure ionic compound with nominal charge states of +1 and -1. In this case, the electronic structure of LiH should be more similar to its cation or anion sublattice, rather than the superposition of the atomic orbitals. Therefore we can compare its total DOS with that of the (artificial but heuristic) cation or anion sublattice, so that to understand profoundly the interaction between the sublattices and how it modifies the charge distribution and electronic structures. It is necessary to point out that this is just to view the same problem from alternative perspective, and is complementary to the traditional decomposition of DOS into atom-centered spherical orbitals. For this reason, in the below we will analyse the electronic structures of LiH by both methods.

LiH at 0 GPa is assumed to be a pure ionic compound formed by sublattices of H$^-$ and Li$^+$. In terms of atomic orbitals, the 2$s$ electron of Li atom is transferred to H-1$s$ state. The left 1$s$ shell in Li$^+$ is thus closed and tightly bound to lithium nucleus, which is rigid and almost unresponsive to atomic environmental changes (see Fig. 1). Therefore the highest occupied valence band in LiH should be the 1$s$ state contributed by H$^-$ sublattice, and the lowest unoccupied conduction
band could be 2$s$ or 2$p$ states from the Li$^+$ sublattice, depending on their relative shift
by local environment. However, since the electrons in H$^-$ are spread out and not tightly bound, one
may argue that at high pressures there might have some overlapping of the wavefunctions between neighboring H$^-$ anions,
leading to bonding $\sigma$ and anti-bonding $\sigma^{\ast}$ states. The latter might become
the lowest unoccupied conduction band, and determine the size of the energy band gap. This simple picture
seems qualitatively reasonable, but its validity needs further confirmation. If decomposing the LiH crystal into separate H$^-$ and Li$^+$ sublattices, and the interaction between them, our calculation predicts that the gap in the H$^-$ sublattice of B1 structure is opened by 1$s$ and 2$p$ states from 0 to 300 GPa, rather than $\sigma$ and $\sigma^{\ast}$ states. The observed little hybridization in Figs. 2 and 3 indicates that the wavefunction overlapping in the H$^-$ sublattice is very small. For the Li$^+$ sublattice, we observe a stronger hybridization between 2$s$ and 2$p$ states, and results in a gap within the conduction band in the total DOS. It should be noted that the 1$s$ state of lithium lies at a much lower energy and thus not shown in Figs. 2-4, in which the $s$ state in the Li$^+$ sublattice refers to Li-2$s$. In the real LiH crystal, the strong interaction
between H$^-$ and Li$^+$ sublattices transfers some electrons from H$^-$ sublattice
back to the Li$^+$ sublattice, and leads to a significant hybridization among H-$1s$, Li-$2s$ and Li-$2p$ in the valence band. The band gap in LiH is almost the same size as in the (artificial) H$^-$ sublattice when at low pressures. But there are two differences: (1) the sublattices interaction now leads to a strong hybridization of $spd$ orbitals of H and Li atoms, and (2) the unoccupied $p$ and
$d$ orbitals in H$^-$ sublattice are greatly depressed by sublattice interactions, and the
gap in LiH is opened between H-1$s$ and Li-2$p$ states. Namely, our calculation at 0 GPa suggests that the top of the
valence band of LiH is dominated by anion 1$s$ state, whereas the bottom of the conduction band is
mainly cation 2$p$ state.
\begin{figure}
\includegraphics{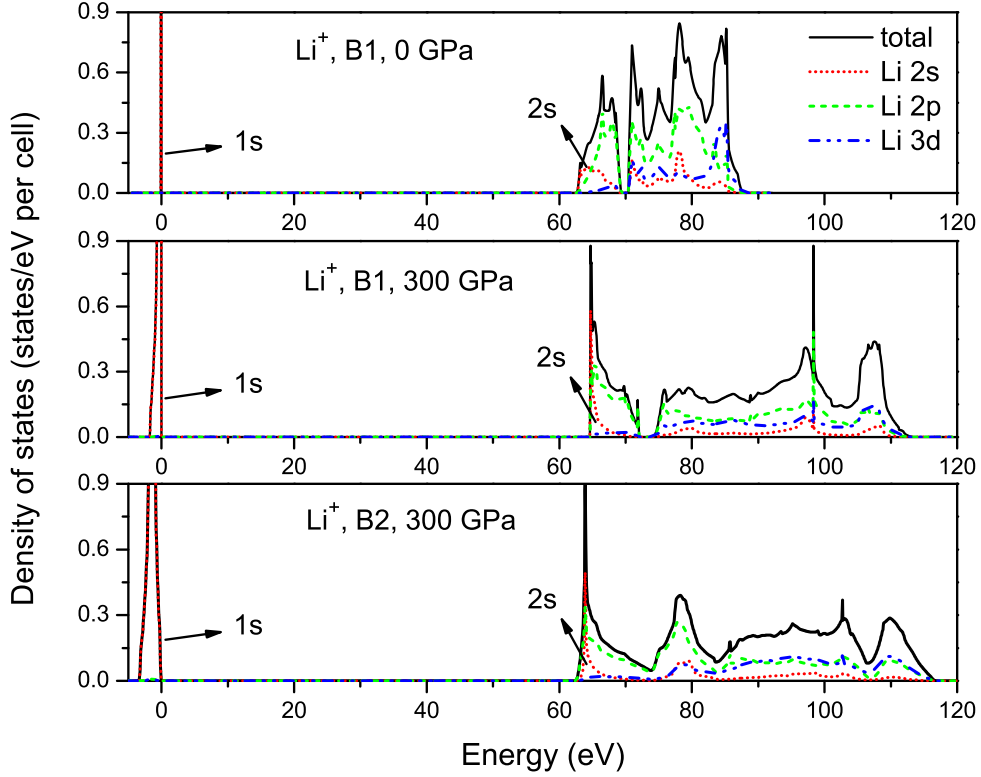}
\caption{(Color online) The total and projected density of states calculated by GW method for Li$^+$ sublattice in B1 structure at 0 and 300 GPa, and in B2 structure at 300 GPa, respectively. The Fermi-level is at zero.}
\end{figure}

\begin{figure}
\includegraphics{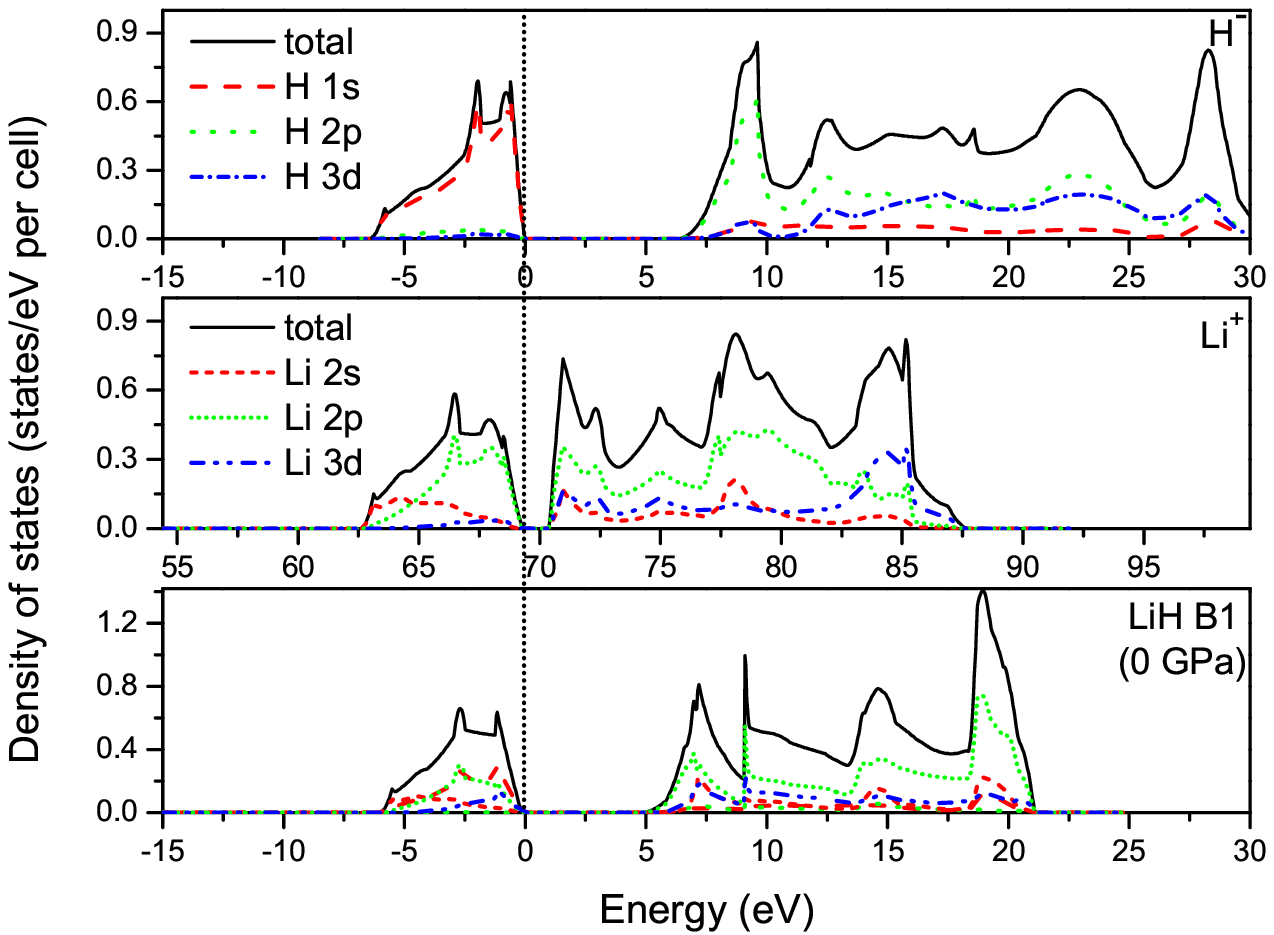}
\caption{(Color online) The total and projected density of states of H$^-$ and Li$^+$
sublattices, and LiH in B1 structure at 0 GPa calculated by GW method, respectively.
The Fermi-level is at zero. Note that the dotted vertical line for Li$^+$ sublattice does not correspond to the Fermi-level, rather it denotes a gap presenting in the conduction band.}
\end{figure}

\begin{figure}
\includegraphics{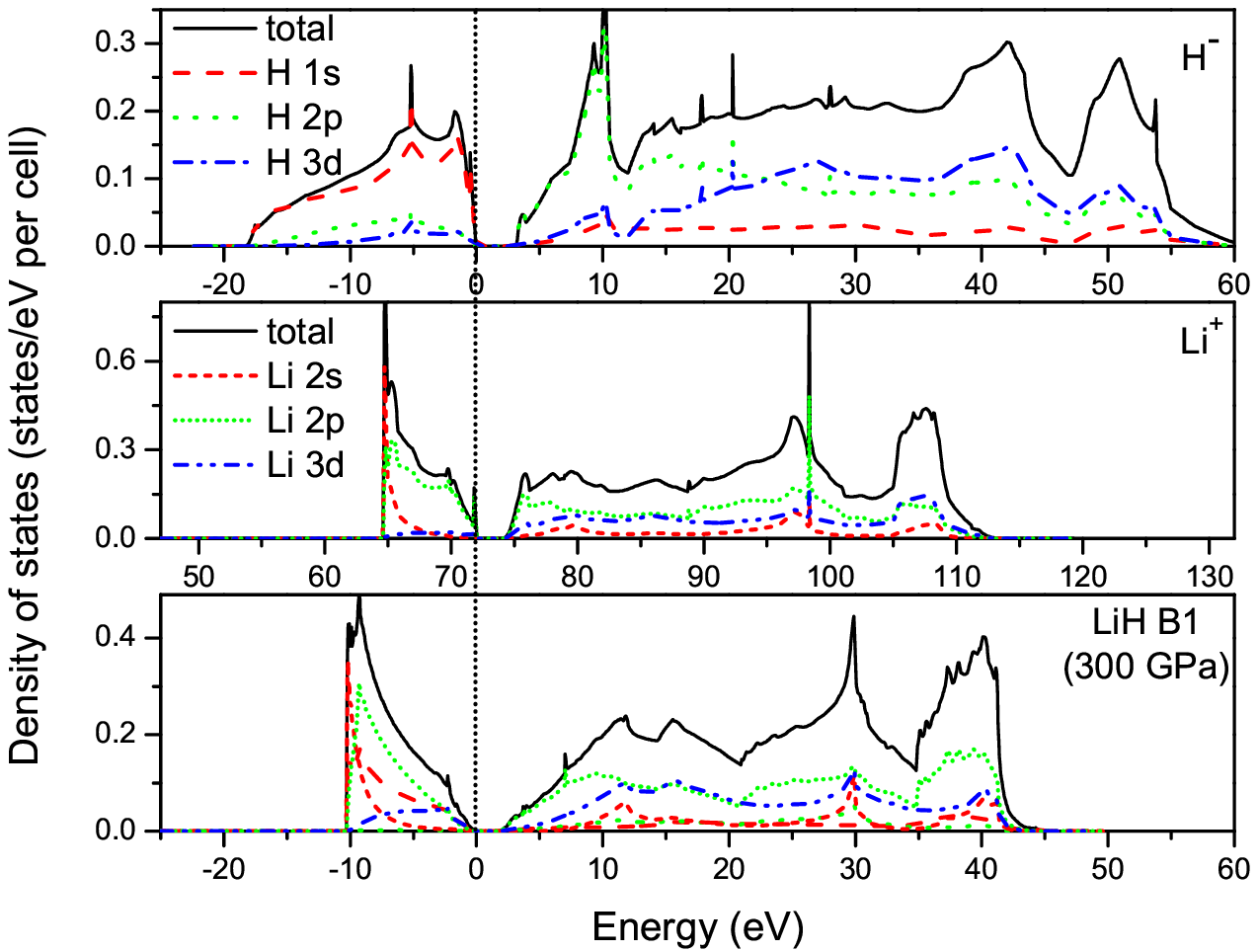}
\caption{(Color online) The total and projected density of states of H$^-$ and Li$^+$
sublattices, and LiH in B1 structure at 300 GPa calculated by GW method, respectively. The
Fermi-level is at zero. Note that the dotted vertical line for Li$^+$ sublattice does not correspond to the Fermi-level, rather it denotes a gap presenting in the conduction band.}
\end{figure}

\begin{figure}
\includegraphics{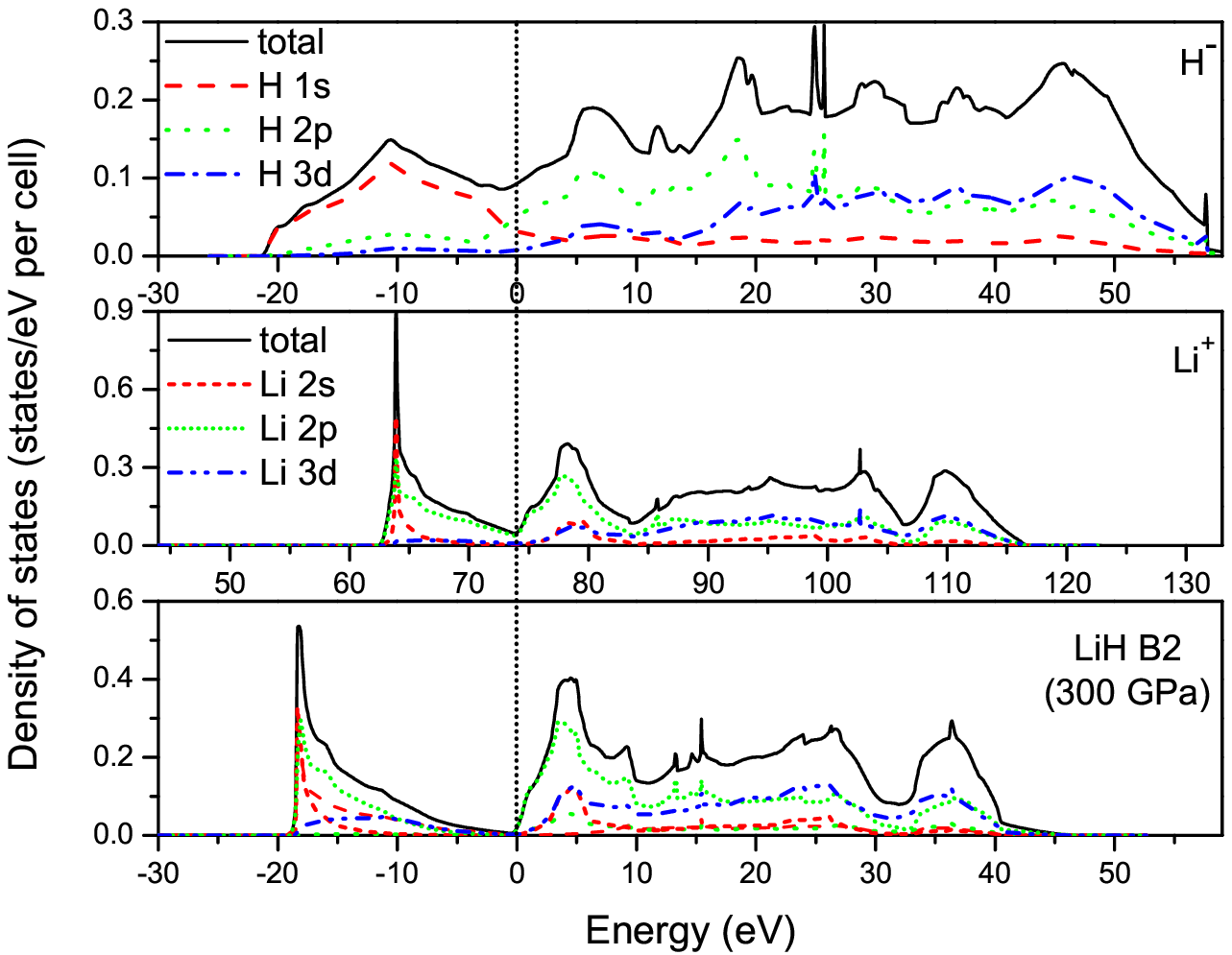}
\caption{(Color online) The total and projected density of states of H$^-$ and Li$^+$
sublattices, and LiH in B2 structure at 300 GPa calculated by GW method, respectively. The
Fermi-level is at zero. Note that the dotted vertical line for Li$^+$ sublattice does not correspond to the Fermi-level, rather it denotes a pseudogap presenting in the conduction band.}
\end{figure}

At higher pressures, taking the B1 and B2 phases at 300 GPa for example, though the sublattice
interactions also transfer electrons back to the Li$^+$ sublattice, the feature of DOS near the
Fermi level is now mainly determined by the Li$^+$ sublattice, rather than by H$^-$ sublattice as shown in Fig. 2.
This is evident from Figs. 3 and 4, which provide the total and projected DOS of separate
H$^-$ and Li$^+$ sublattices and the real crystal LiH, respectively. It also can be
found that the stability of B2 phase with respect to B1 phase mainly comes from the larger valence band
width of the former (18.91 eV versus 10.39 eV), i.e., the delocalization of the valence states.
For the B1 phase at 300 GPa, GW pushes Li-2$p$ orbital away from the Fermi level, and creates a gap of 2.05 eV,
whereas in B2 phase GW broadens the valence band width to 18.91 eV, as illustrated in Fig. 4.
Therefore the band gap in B1 phase at high pressures is opened between cation 2$p$
state and hybridized $spd$ state. By comparing the change of the total and projected DOS
of LiH at around the B1 $\rightarrow$ B2 transition pressure of 300 GPa with that at 0 GPa calculated by GW method,
it is evident that compression delocalizes H-1$s$, Li-2$s$ and Li-2$p$ states. This leads to a strong
hybridization among them and broadens the valence band and conduction band width,
thus reduces the band gap. The phase transition to B2 phase at 300 GPa broadens the valence
band width greatly, and for this reason the band gap closes up. At the same time the conduction
band width is slightly narrowed, with unoccupied Li-2$p$ state localized just above the Fermi level.

\subsection{Vibrational free energy at high pressures and temperatures}
Although the GW mehod provides a better description of the electronic structures, it is computationally demanding and does not lead to a better total energy and forces. At the LDA or GGA level, the total energy and forces are usually well produced. For this reason, GGA is used to calculate the total energies in this work, which are then fitted to the Vinet EOS. The fitted parameters of LiH at the considered pressures range are shown in Table \uppercase\expandafter{\romannumeral1}. It is necessary to point out that we fit the Vinet EOS to different pressure segments for different purpose separately. For those of high pressure fittings, their parameters do not have any physical implications. Only those of B1 phase fitted to data in 0-100 GPa can be compared to the experimental data directly. For lattice dynamics, the first-principles QHA is employed to calculate phonon spectra. In order to represent the discrete vibrational free energy data accurately, the double-Debye model (2DM) is employed. Variation of the Debye temperatures as a function of volume is described by the Gr\"{u}neisen parameter. Table \uppercase\expandafter{\romannumeral2} lists the fitted parameters of 2DM for both B1 and B2 phases of $^6$LiH, $^6$LiD, and $^6$LiT, as well as the $^7$LiH and $^7$LiD in B1 phase. In the 2DM, it is required to satisfy a condition of $\theta_{A}<\theta_{0}<\theta_{B}$.
If $\theta_{A}$ = $\theta_{B}$ = $\theta_{0}$, it reduces back to the 1DM. Hence the deviation of $\theta_{A(B)}$ from $\theta_{0}$  measures the significance of 2DM against 1DM.

\begin{table}
\begin{threeparttable}[b]
\caption{Fitted Vinet EOS parameters $E_0$, $K_0$, $K^{'}{_0}$ and $V_0$ for lithium
hydride, the reference state is at the lowest pressure for each considered pressure segment.}
\begin{ruledtabular}
\begin{tabular}{ccccc}
& $E_0$ (eV)  &	$K_0$ (GPa) &	$K^{'}{_0}$ & $V_0$ ({\AA}$^{3}$)  \\
\hline
B1 (100 GPa-450GPa) &	-8.34 &	1.72  &	6.32 &	30.89 \\
B2 (100 GPa-450GPa) &	-7.81 &	0.26  &	7.64 &	42.92 \\
B1 (0 GPa-450 GPa)  &	-7.92 &	24.18 &	4.39 &	16.92 \\
B1 (0 GPa-100 GPa)  &	-7.88 &	33.63 &	3.81 &	16.14 \\
Expt.              &	---      &	34.24\tnote{e}&	3.80$\pm$0.15\tnote{e} &16.72\tnote{f} \\
\end{tabular}
\footnotesize
\begin{tablenotes}
  \item[e]Reference [\onlinecite{Vidal131}].\item[f]Reference [\onlinecite{Gerlich1587}].
  \end{tablenotes}
\end{ruledtabular}
\end{threeparttable}
\end{table}

\begin{table}
\caption{Parameters of the double-Debye model obtained by fitting to first-principles phDOS
of both B1 and B2 phases of $^6$LiH, $^6$LiD, $^6$LiT and $^7$LiT within a pressure range
from 100 to 450 GPa. Also given are the B1 phase of $^7$LiH and $^7$LiD from 0 to 100 GPa.
The reference state of ${{\theta }_{\left\{ 0,\text{ }A,\text{ }B \right\}}}^{0}$ is at the highest
pressure end for each considered pressure range.}
\begin{ruledtabular}
\begin{tabular}{cccccccccc}
&${{\theta }_{0}}^{0}$ (K)&	${{\alpha }_{0}}$ &	${{\beta }_{0}}$ ({\AA}$^{-3}$) &	${{\theta }_{A}}^{0}$ (K)&	${{\alpha }_{A}}$
&	${{\beta }_{A}}$ ({\AA}$^{-3}$) &	${{\theta }_{B}}^{0}$ (K) &	${{\alpha }_{B}}$&	${{\beta }_{B}}$ ({\AA}$^{-3}$) \\
\hline
$^6$LiH(B1) &	3036.65 &	0.326  &	0.053 &	1474.33 &	-0.462 &	0.138  &	4383.04 &	0.506 &	0.037  \\
$^6$LiD(B1) &	2553.45 &	0.318  &	0.057 &	1316.22 &	-0.606 &	0.125  &	3229.39 &	0.593 &	0.027  \\
$^6$LiT(B1) &	2307.49 &	0.319  &	0.057 &	1222.27 &	-0.098 &	-0.005 &	2809.79 &	0.807 &	-0.010 \\
$^7$LiT(B1) &   2220.26	&   0.320  &     0.057 &---&---&---&---&---& ---\\
$^6$LiH(B2) &	2995.69 &	-0.286 &    0.218 & 1844.87 &	0.298  &	0.118  &	4205.52 &	0.802 &	-0.004 \\
$^6$LiD(B2) &	2518.94 &	0.249  &	0.112 &	1974.72 &	-0.887 &	0.405  &	3139.34 &	0.844 &	0.010  \\
$^6$LiT(B2) &	2276.23 &	0.252  &	0.111 &	2153.76 &	-3.779 &	1.024  &	3245.11 &	1.351 &	-0.004 \\
$^7$LiT(B2) &	2423.82 &	0.249  &	0.112 &	1718.53 &	-2.396 &	0.737  &	3066.68 &	1.023 &	0.007  \\
$^7$LiH(B1)	&   2038.74	&   0.420  &    0.039 &1137.90	&   0.096  &    0.057  &    2926.23	&   0.714 & 0.008 \\
$^7$LiD(B1)	&   1714.40	&   0.420  &    0.039 &1146.12	&   -0.388 &    0.087  &    2103.34	&   0.855 & -0.010 \\
$^6$LiH(B1)$^{a}$  &	3036.65 &	0.404  &	0.041 &	1474.33 &	-0.116 &	0.076  &	4383.04 &	0.616 &	0.018 \\
\end{tabular}
\footnotesize
\begin{tablenotes}
  \item[a]$^a$Fitting to the B1 phase of $^6$LiH from 0 to 450 GPa.
  \end{tablenotes}
\end{ruledtabular}
\end{table}

For comparison, the single-Debye model is also assessed. Table \uppercase\expandafter{\romannumeral3} compares the relative errors of 1DM and 2DM when used to reproduce the QHA vibrational free energy in both B1 and B2 structures for $^6$LiH, $^6$LiD, $^6$LiT, and $^7$LiT at different pressures and temperatures. It can be seen that at 300 K, the vibrational free energy of $^6$LiH reproduced by
1DM has the largest relative errors ranging from 7.27\% to 10.98\% for both B1 and B2 phases at 100 and 450 GPa, respectively. Note that the relative errors at 3000 K are greatly reduced. Heavier hydrogen isotopes have less relative errors. This is because that the large mass of hydrogen isotopes reduces its vibrational frequencies, and the phonon spectrum becomes more similiar to that of a single-Debye model. When the 2DM is employed to represent the QHA results, all of the relative errors in $^6$LiH, $^6$LiD, and $^6$LiT are reduced by one or two orders in the magnitude. Note that for the B2 structure at 100 GPa and 3000 K, the relative error of 2DM is slightly smaller than that of 1DM, and both are less than 3.97\%. This indicates that 1DM and 2DM have similar precision to reproduce the QHA data under this condition, but 2DM is much better in all other cases. The good performance of 2DM is attributed to the multiple peaks in the phonon spectrum of lithium hydrides, which are easier to be captured by 2DM than 1DM. Figure 5 displays the phDOS of the first-principles QHA and that of 2DM in B1 and B2 structures of $^6$LiH at 100 and 450 GPa, respectively. It is evident that the shape of $g_{FP}(\omega)$ is more close to 2DM than 1DM, the latter has just one peak.

For the standard single-Debye model, the thermal pressure is given by:
\begin{equation}\label{15}
{{P}_{th}}=\frac{{{\gamma }_{0}}^{{}}}{V}(\frac{9}{8}{{k}_{B}}{{\theta }_{0}}+3{{k}_{B}}TD(\frac{{{\theta }_{0}}}{T})).
\end{equation}
Here the Gr\"{u}neisen parameter $\gamma_{0}$ is a smooth function of volume, and usually can be described by Eq. (\ref{lntheta0AB}) approximately. The Gr\"{u}neisen parameters $\gamma_{A(B)}$ in 2DM also have a similar behavior as $\gamma_{0}$, and Eq. (\ref{lntheta0AB}) works well for most cases, except for $^7$LiT. In Supplementary Figs. S1 and S2, we plot the variation of $\theta_{A}$ and $\theta_{B}$ of $^7$LiT in B1 and B2 phases as a function of volume. It can be seen that both $\theta_{A}$ and $\theta_{B}$ are well-behaved in B2 phase, and can be described by Eq. (\ref{lntheta0AB}) very well. However, $\theta_{A}$ and $\theta_{B}$ show irregular variation in B1 phase, and the corresponding Gr\"{u}neisen parameters thus can not be derived. The exact cause of this abnormality is unclear, and requires further investigation in the future. Therefore, for the B1 phase of $^7$LiT we simply use the 1DM to reproduce the QHA results. This gives a slightly larger error at low temperatures, as shown in Table \uppercase\expandafter{\romannumeral3}.

\begin{table}
\caption{The relative error of the vibrational free energy calculated by single-Debye model and
double-Debye model with respect to the first-principles QHA in lithium hydrides at about 100 and 450 GPa, respectively.}
\begin{ruledtabular}
\begin{tabular}{ccccccccccc}
\multicolumn{3}{c}{\multirow{2}{*}{Error(\%)}}          & \multicolumn{4}{c}{single-Debye model}& \multicolumn{4}{c}{double-Debye model}\\
\cline{4-7}
\cline{8-11}
\multicolumn{3}{c}{}                                    & $^6$LiH &	$^6$LiD & $^6$LiT &	$^7$LiT & $^6$LiH &	$^6$LiD & $^6$LiT &	$^7$LiT \\
\hline
\multirow{4}{*}{B1}	&\multirow{2}{*}{100 GPa}	&300 K  &	7.33  &	2.75    &	1.33  &	5.21    &	0.05  &	0.02    &	0.09  &  ---	    \\
\multirow{4}{*}{}	&\multirow{2}{*}{}          &3000 K &	0.60  &	0.18    &	0.09  &	0.45    &	0.009 &	0.004   &	0.04  &  ---	    \\
\multirow{4}{*}{}   &\multirow{2}{*}{450 GPa}	&300 K  &	10.60 &	6.18    &	4.92  &	1.71    &	0.02  &	0.01    &	0.003 &  ---	    \\
\multirow{4}{*}{}   &\multirow{2}{*}{}          &3000 K &	4.04  &	0.94    &	0.48  &	0.05    &	0.01  &	0.003   &	0.02  &  ---	    \\
\multirow{4}{*}{B2}	&\multirow{2}{*}{100 GPa}	&300 K  &	10.98 &	3.95    &	2.51  &	2.87    &	2.60  &	0.14    &	0.61  &	0.26    \\
\multirow{4}{*}{}	&\multirow{2}{*}{}          &3000 K &	3.97  &	0.23    &	0.14  &	0.15    &	3.50  &	0.11    &	0.10  &	0.09    \\
\multirow{4}{*}{}   &\multirow{2}{*}{450 GPa}	&300 K  &	7.27  &	2.56    &	1.08  &	1.52    &	0.03  &	0.03    &	0.03  &	0.03    \\
\multirow{4}{*}{}   &\multirow{2}{*}{}          &3000 K &	2.40  &	0.31    &	0.05  &	0.09    &	0.14  &	0.07    &	0.07  &	0.05    \\
\end{tabular}
\end{ruledtabular}
\end{table}

\begin{figure}
\includegraphics{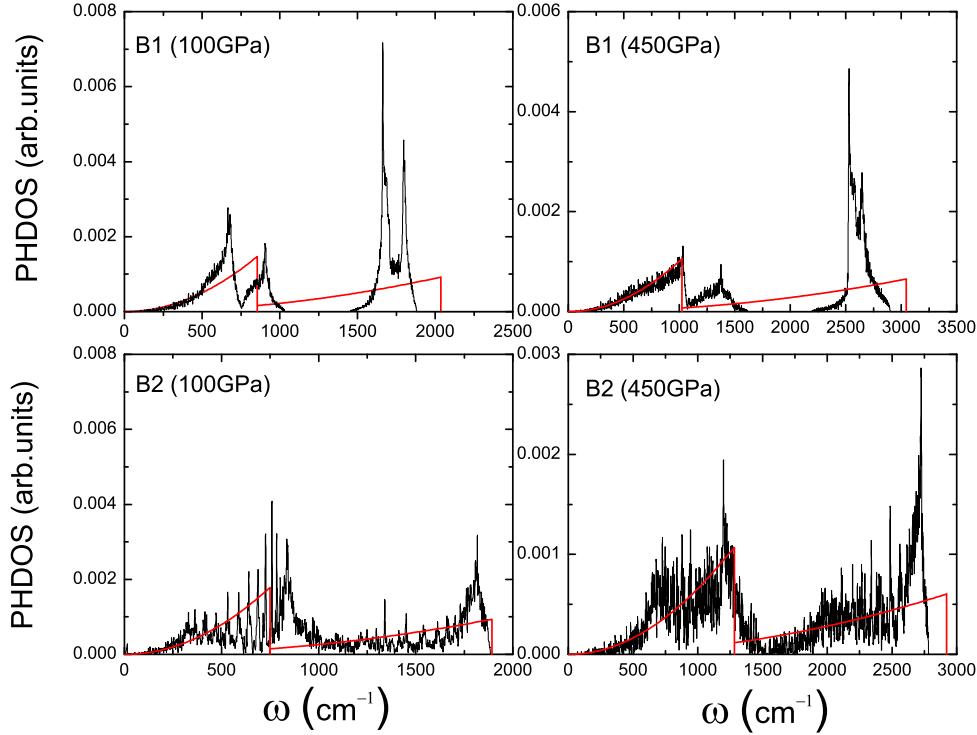}
\caption{(Color online) The phonon density of state (phDOS) of $^6$LiH in B1 and B2
structures at 100 and 450 GPa, respectively. The black lines denote the first-principles QHA data, and red lines are for 2DM.}
\end{figure}

\begin{figure}
\includegraphics{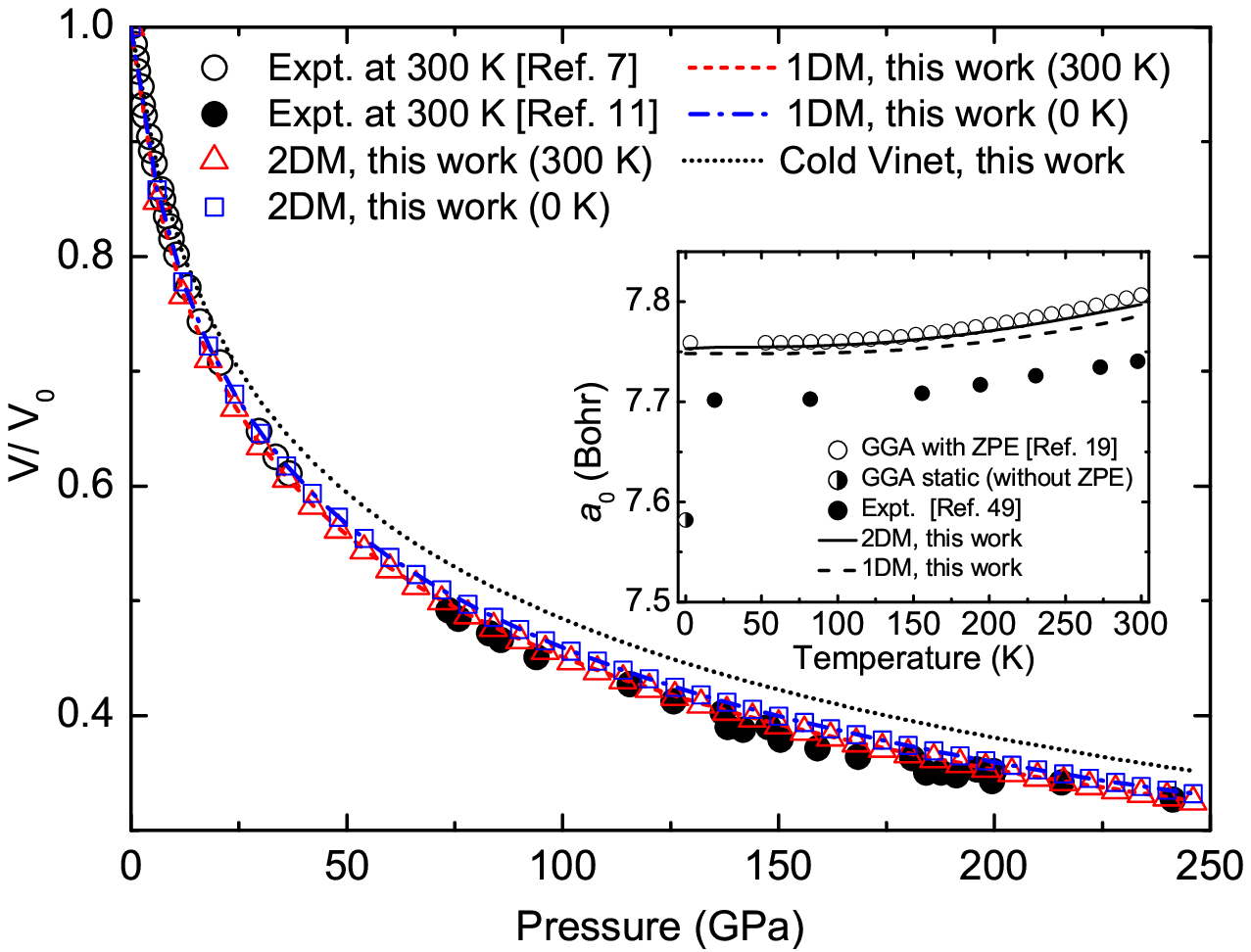}
\caption{(Color online) Comparison of the static equation of state for solid $^7$LiH obtained by different methods, V$_0$ is the specific
volume at ambient conditions.
Inset: Variations of the lattice constant a$_0$ with temperature for the B1 phase of $^7$LiH at ambient pressure.}
\end{figure}

\subsection{Equation of state and B1-B2 solid phase boundary}
\begin{figure}
\includegraphics{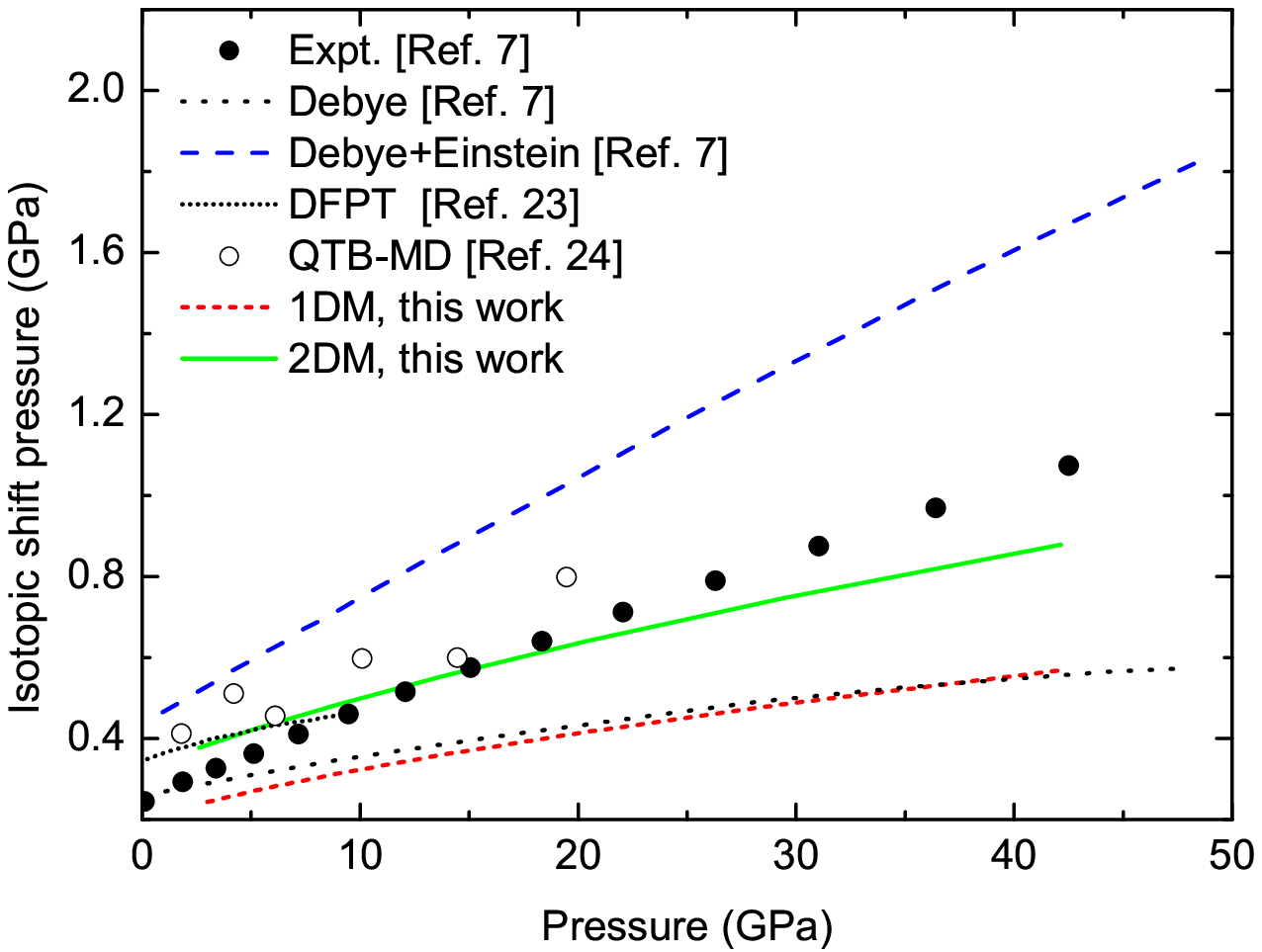}
\caption{(Color online) Isotopic shift in pressure (i.e., the pressure difference
between $^7$LiH and $^7$LiD at a given volume) as a function of the pressure of $^7$LiD at 300 K.}
\end{figure}

Due to the high accuracy of 2DM in reproducing the vibrational free energy, it can be used to calculate the EOS of lithium hydrides. Previous theoretical calculations \cite{Mukherjee403, Yu086209} showed a slight deviation from the experimental EOS data \cite{Lazicki054103}. This was considered as due to the neglect of ZPE. Figure 6 displays our EOS of $^7$LiH at 300 K obtained with different methods by comparison to the static DAC experimental data of Loubeyre $et$ $al.$ \cite{Loubeyre10403} and Lazicki $et$ $al$ \cite{Lazicki054103}. Note that in this figure, the cold pressure curve of Vinet EOS does not include the ZPE contribution, whereas those marked as 0 K do include ZPE. It can be seen that the cold EOS (represented by Vinet EOS model) is remarkably incompressible than the experimental data, especially at high pressures. The vibrational contribution softens the EOS greatly. In particular, the 300 K isotherm is in good agreement with the experimental data, whereas the 0 K isotherm is still less compressible. This reveals that zero-point motion and temperature play a significant role in the EOS of lithium hydrides. The change of the lattice constant with temperature for the B1 phase in $^7$LiH at ambient pressure is shown in the inset of Fig. 6. It can be seen that our EOS is in good agreement with other theoretical \cite{Yu086209} and experimental data \cite{Smith246}. It should be noted that both 2DM and 1DM give a similiar static compression curve in the whole considered pressure range.

The isotopic shift in the pressure between $^7$LiH and $^7$LiD at 300 K has been measured experimentally \cite{Loubeyre10403}. Previous DFPT results \cite{Roma203} revealed the important role of ZPE in this isotopic shift at low pressures. Here we employ the 2DM and 1DM to calculate the isotopic shift for the whole pressure range considered in the experiment. As shown in Fig. 7, our 1DM results are in good agreement with the previous results that also used the standard single-Debye approximation. Both results underestimate the isotopic shift. On the other hand, the mixed Debye-Einstein (with the transverse optical phonons represented by Einstein model) overestimates this isotopic shift. However, when 2DM is used to represent the QHA data, the isotopic shift in pressure is accurately reproduced. This suggests that it is the function form of the single-Debye model that deteriorates the QHA results. At high pressures, the slight deviation between 2DM results and the experimental data might be due to the anharmonicity that was not included in our first-principles QHA calculations, as implied by the QTB-MD \cite{Dammak435402} calculations.

Moreover, we also investigate the isotopic effects on the B1-B2 solid phase boundaries of lithium hydrides. Previous theoretical studies mainly focused on this transition of $^7$LiH at 0 and 300 K, and estimated a transition pressure spanning from 200 to 500 GPa \cite{Hochheimer139, Ghandehari2264, Lazicki054103, Hammerberg617, Martins7883, Zhang104115, Mukherjee103515, Wang470}. The finite temperature phase transition and isotopic effects, however, are not explored. The calculated B1-B2 solid phase boundaries of $^6$LiH, $^6$LiD, $^6$LiT, and $^7$LiT are displayed in Fig. 8. Those of $^7$LiH and $^7$LiD are not listed because their isotopic effect is very close to $^6$LiH and $^6$LiD, respectively. 2DM is employed to calculate the phase boundaries for $^6$LiH, $^6$LiD, and $^6$LiT, whereas 1DM is used for $^7$LiT because the double-Debye cannot be well defined for this isotope when in the B1 phase. The inset of Fig. 8 demonstrates the relative errors of 1DM  with respect to that of 2DM in the B1-B2 phase transition pressures of $^6$LiH, $^6$LiD, and $^6$LiT when temperature varying from 0 to 3000 K. It can be seen that the relative errors decrease with increasing temperature, and the largest relative error is about 5\% in $^6$LiH at 0 K. The magnitude of these errors cannot be ignored when describing the B1-B2 solid phase boundary.

\begin{figure}
\includegraphics{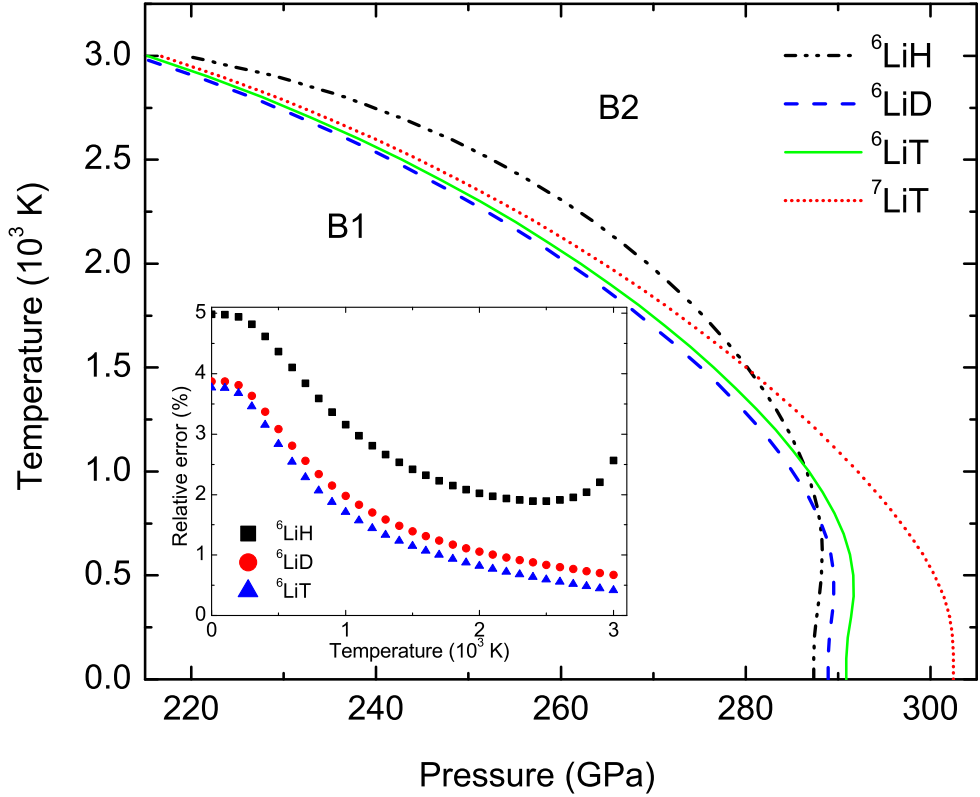}
\caption{(Color online) The B1-B2 solid phase boundaries of $^6$LiH, $^6$LiD and $^6$LiT calculated by double-Debye model, and that of $^7$LiT by single-Debye model, respectively. Inset: the relative error in B1-B2 transition pressure of 1DM against 2DM.}
\end{figure}

From Fig. 8, it is evident that the isotopic effects on the B1-B2 phase boundaries are striking. At high temperatures above 2000 K, there are remarkable isotopic shift between $^6$LiH and $^6$LiD. When temperature descreases, the isotopic effect between $^6$LiT and $^7$LiT also becomes large. But this might be due to the errors in $^7$LiT, because its phase boundary is calculated with 1DM. At 0 K, the transition pressure difference between $^6$LiH and $^7$LiT is about 15 GPa. With increasing temperature this difference reduces, and finally overturns at 1490 K and 280 GPa. Beyond that temperature, $^6$LiH has higher B1-B2 transition pressure than $^7$LiT. Note that $^6$LiH also reverses the relative position of its boundary with respect to $^6$LiD and $^6$LiT. At very high temperatures of close to 3000 K, the isotopic shift diminishes. It is necessary to point out that except $^7$LiT, all of $^6$LiX (X = H, D, T) show a weak reentrant feature in their B1-B2 phase boundary. Namely, within a narrow pressure range
just above the 0 K transition pressure, increasing temperature will transform the compound back to
the B1 phase, and further increasing temperature will bring it back to the B2 phase again.
By far, it is unclear whether it is a unique property of lithium hydrides or also shared by other alkali hydrides.

\subsection{Phase diagram}

With above comprehensive calculations and analysis, we finally reach the stage to construct a finite temperature phase diagram for LiH. This phase diagram is fundamental to understand the high-pressure and high-temperature thermodynamics of lithium hydrides. Combining our calculated B1-B2 finite temperature phase boundary of $^6$LiH, and the previously calculated melting curve of B1
phase for $^7$LiH reported by Ogitsu $et$ $al$ \cite{Ogitsu175502}, as well as its extrapolation using Kechin
equation \cite{Kechin052102} of ${{T}_{m}}(P)=790{{[1+0.3911(P+0.28)]}^{0.3221}}{{e}^{-0.001373(P+0.28)}}$,
where ${{T}_{m}}(P)$ is the melting temperature at a given pressure $P$, we obtain a first-principles
phase diagram of LiH, and show it in Fig. 9. The B1-B2-liquid triple point is determined at 241 GPa and 2413 K.
Here we note that the isotopic effects on the melting curves between $^6$LiH and $^7$LiH is negligible,
because of the small relative mass difference between them.

\begin{figure}
\includegraphics{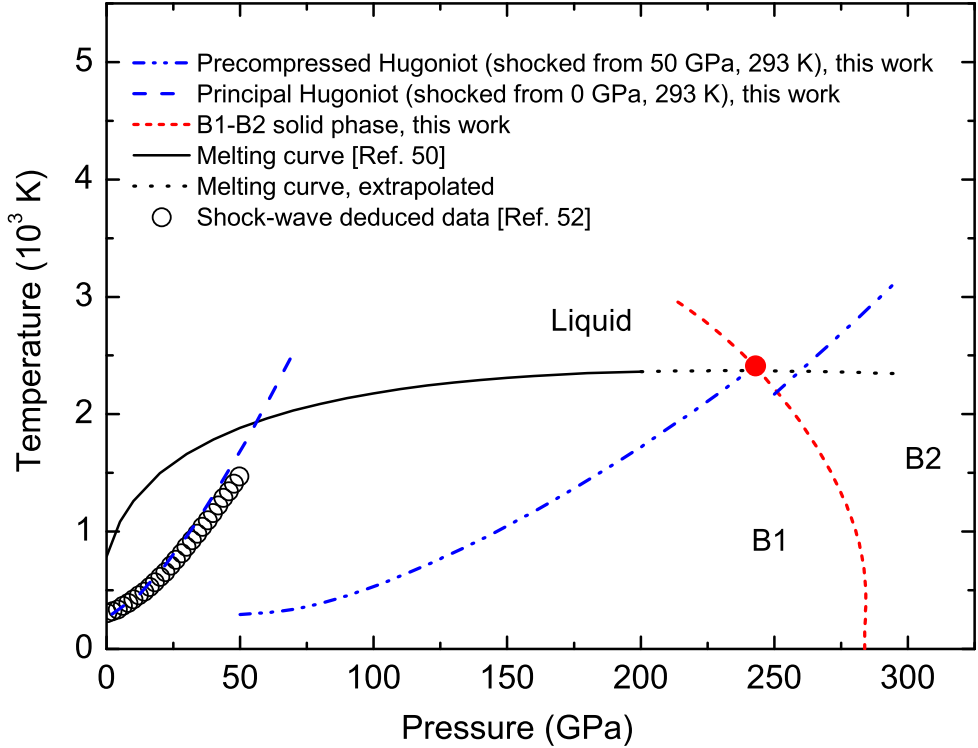}
\caption{(Color online) Calculated phase diagram of $^6$LiH. The short dashed line and short dotted line are
the B1-B2 solid phase boundary and the extrapolation of the previous melting curve, respectively.
The dashed line is the calculated principal Hugoniot with the shock melting point at 56 GPa
and 1923 K, and the dash-dot-dotted line is the Hugoniot with precompression and shocked from 50 GPa and 293 K,
which passes through the triple point (red solid circle) at around 241 GPa and 2413 K.}
\end{figure}

Besides the static experiment such as DAC, dynamical compression is also an important method to explore the high pressure physics.
The principal shock Hugoniot of $^6$LiH calculated by 2DM is shown in Fig. 9, and is compared with the deduced data of the shock-wave experiment reported by Marsh \cite{Marsh1972} (the details of calculating shock Hugoniots from first-principles calculations are referenced to Ref. [\onlinecite{geng2005shock}]). It can be seen that our results are in good accordance with the deduced data of shock wave experiments. Only when the shock pressure is higher than 25 GPa, our 2DM predicts a slightly higher shock temperature. However, the slight deviation in shock temperature might not be due to the QHA data or the fitting error in 2DM. It should be noted that the samples of the shock wave experiment in Ref. [\onlinecite{Marsh1972}]) contained a little impurities (4.5\% $^7$Li). This might modify the phonon spectra and lattice specific heat, and thus reduce the lattice dynamics contributions.

Our calculation predicts that the shock melting occurs at 1923 K and 56 GPa, which is
far from the stable region of B2 solid phase. As shown in Fig. 9, a direct shock of LiH
cannot cross the B1-B2 phase boundary. Besides isentropic or multiple shock compression
techniques, precompression of the sample at low pressure is an alternative route to enter the B2 solid phase. We find that in order to pass through the triple point and to enter the B2 phase, it requires at least a precompression of 50 GPa at 293 K. The resultant precompression plus shock Hugoniot is also shown in Fig. 9. When entering the B2 phase along this path, there is a temperature drop of 230 K. The corresponding volume collapse is about 4.6\% (see Fig. S3 in the supplemental material). By comparison, the same B1 $\rightarrow$ B2 transition at 0 K has only 1.2\% volume collapse.

\section{CONCLUSIONS}
In summary, we have performed comprehensive first-principles calculations to
understand the electronic structures, thermodynamic properties, and phase diagram of lithium hydrides.
By investigating the electronic structures, we found that LiH is not a pure ionic compound with nominal charge state. There is a strong interaction between Li$^+$ and H$^-$ sublattices of the assumed pure ionic compound, which leads to a charge transfer from the latter back to the former, and results in a strong $spd$ hybridization in the real LiH. At low pressures, the electronic structure near the Fermi level is determined by H$^-$ sublattice, whereas it is dominated by Li$^+$ sublattice at high pressures.
The first-principles QHA was used to describe the lattice dynamics. The discrete phonon data were then fitted to a
double-Debye model with only nine parameters, which accurately reproduces the first-principles vibrational free energy. The isotopic effects on the equation of states and the B1-B2 solid phase boundaries of lithium hydrides are also well modelled. Furthermore, the phase diagram of LiH was amended and completed by first-principle method, which predicts a triple point at 241 GPa and 2413 K. Extended analysis revealed that a precompression of the sample to 50 GPa will make the shock Hugoniot go through the B1-B2 boundary and enter the B2 solid phase with a discontinuity having large volume collapse. Considering that lithium hydrides are applied widely in industry and nuclear power engineering, our results will be practical helpful and stimulate further theoretical and experimental investigations.

\section*{ACKNOWLEDGMENTS}
This work is supported by the National Natural Science Foundation of China
under Grant Nos. 11274281 and 11174214, the CAEP Research Projects under
Grant Nos. 2012A0101001 and 2015B0101005, and the NSAF under Grant No. U1430117.

\bibliography{Reference}
\end{document}